\documentclass[%
prapplied,
superscriptaddress,
floatfix,
citeautoscript,
 amsmath,amssymb,
 reprint,%
]{revtex4-2}

\makeatletter 
\renewcommand\NAT@biblabelnum[1]{#1\,.}
\makeatother

\usepackage{graphicx}
\usepackage{epstopdf}
\usepackage{dcolumn}
\usepackage{xcolor}
\usepackage{wasysym}
\usepackage{hyperref}
\usepackage{cleveref}
\usepackage{multirow}

\graphicspath{{./graphics/}}
\definecolor{tred}{rgb}{0.71,0.16,0.43}

\begin{document}

\title{Tailoring potentials by simulation-aided design of gate layouts for spin qubit applications} %Title of paper

\author{Inga Seidler}
\author{Malte Neul}
\author{Eugen Kammerloher}
\author{Matthias Künne}
\affiliation{JARA-FIT Institute for Quantum Information, RWTH Aachen University, 52074~Aachen, Germany}
\author{Andreas Schmidbauer}
\author{Laura Diebel}
\affiliation{Fakultät für Physik, Universität Regensburg, 93040~Regensburg, Germany}
\author{Arne Ludwig}
\author{Julian Ritzmann}
\author{Andreas D. Wieck}
\affiliation{Applied Solid State Physics, Ruhr-Universität Bochum, 44801~Bochum, Germany}
\author{Dominique Bougeard}
\affiliation{Fakultät für Physik, Universität Regensburg, 93040~Regensburg, Germany}
\author{Hendrik Bluhm}
\author{Lars R. Schreiber}
\email{lars.schreiber@physik.rwth-aachen.de}
\affiliation{JARA-FIT Institute for Quantum Information, RWTH Aachen University, 52074~Aachen, Germany}

\date{\today}
\definecolor{Blue}{rgb}{0,0,1}
\definecolor{Green}{rgb}{0,1,0}
\definecolor{Red}{rgb}{1,0,0}
\newcommand{\LS}[1]{\textcolor{Blue}{\small LS: #1}}
\newcommand{\HB}[1]{\textcolor{Green}{\small HB: #1}}
\newcommand{\MN}[1]{\textcolor{Red}{\small MN: #1}}
\newcommand{\IS}[1]{\textcolor{orange}{\small IS: #1}}
\begin{abstract}
Gate-layouts of spin qubit devices are commonly adapted from previous successful devices. As qubit numbers and the device complexity increase, modelling new device layouts and optimizing for yield and performance becomes necessary. Simulation tools from advanced semiconductor industry need to be adapted for smaller structure sizes and electron numbers. 
Here, we present a general approach for electrostatically modelling new spin qubit device layouts, considering gate voltages, heterostructures, reservoirs and an applied source-drain bias. Exemplified by a specific potential, we study the influence of each parameter. We verify our model by indirectly probing the potential landscape of two design implementations through transport measurements. We use the simulations to identify critical design areas and optimize for robustness with regard to influence and resolution limits of the fabrication process.
\end{abstract}
\maketitle

\section{Introduction}
Demonstrator devices for electron spin qubits have been shown to work with high manipulation \cite{Yoneda2018,Zajac2018,Watson2018,Xue2019,Xue2021,madzik2022,Noiri2022,Philips2022,Mills2022} and readout fidelities \cite{Connors2020,Noiri2020, Kammerloher2021,Mills2022} and indicate a possible path to scaling to a quantum computer \cite{Hollenberg2006,Vandersypen2017,Veldhorst2017,Li2018,Boter19,Boter22}.
Gate-layouts of most demonstrator devices are closely related to or copies of previous devices of a research group or of published layouts. Their functionality has been mostly tested as completed devices and therefore only few iterations are made due to the relatively slow feedback cycle. 
Scaling up to larger qubit numbers requires optimized and new device layouts. As devices become more complex, testing of many device generations is not sufficient to achieve high yield and robustness to material variations. To this end, simulations are needed to predict functionalities and finally increase the feedback cycle from device measurement to fabrication. Simulation tools are extensively used throughout the advanced semiconductor industry for high complexity devices, as for example technology computer-aided design which is used for transistor, photo detector and miniature solar cell designs \cite{tiwari19,jacob05,GREULICH2015}. 

In contrast to these applications, spin qubit devices require the accuracy of single electron control, include tunnel barriers and are mostly based on smaller designs. These simulations require additional quantum mechanical constrains. Adaptions of the TCAD software allowing for quantum mechanical restrictions have been implemented with the focus on optimizing qubit distances for manipulation and tunnel couplings \cite{Mohiyaddin2019}. There are simulation-based micro-magnet designs optimised for fast and precise spin manipulation by electric dipolar spin resonance \cite{pioro-ladriere2008,Neumann2015,Philips2022}. The coupling to electron reservoirs has been calculated by simulating the induced potential in the 2DEG and approximating the Hamiltonian \cite{Klos2018,Rochette2019}.  A detailed comparison to experimentally applied voltages has only been shown for gate pinchoffs \cite{Chatzikyriakou2022}.

These modelling efforts are based on ideal devices layouts. However, the consideration of process constrains promises higher functionality and device yields. To that extend, fabricational variations such as line-edge roughness on the order of a few new nanometers \cite{Miyoshi15}, and the limitations of patterning with electron beam lithography at non-vertical angles need to be considered \cite{RANGELOw94}.

Here, we present a general approach to electrostatic modelling of a targeted potential taking into account the gate structure, doping, reservoirs and applied bias. We consider the influence of fabricational variability to optimize for stability and few device iterations. For a specific targeted potential, implementations for both undoped and doped heterostructures as well as a depletion and accumulation mode designs are considered. For two different heterostructure implementations, the simulated functionality is ratified experimentally.

\section{Target Potential}
Our approach can be used to implement any targeted potential. To exemplify the possible application, we discuss one specific potential to highlight the studied parameters. For this example, the aim is to form a quantum dot (QD) that has identical tunnel couplings to source and drain reservoir, but a significantly larger capacitive coupling to the source than the drain reservoir (Fig. \ref{fig:setup}a). We implement this configuration by forming a QD potential with a sharp tunnel barrier to the source reservoir and the drain reservoir with an added potential section in between the second barrier and the drain reservoir. The potential in this added section, named slide, slowly decreases the potential to the chemical potential of the drain reservoir over a distance of several 100 nm (Fig. \ref{fig:setup}b). This specific potential is studied as a QD charge sensor with enhanced performance \cite{Kammerloher2021}. A double quantum dot (DQD) is added next to it, which can host the spin qubit.  The capacitive coupling between the QD of the sensor and the DQD must be high for good charge sensitivity. 

\section{Gate Layouts}

To highlight the versatility of our simulation approach regarding the heterostructure and the charge accumulation, we choose three different realization, including both a doped and an undoped heterostructure as well as a depletion- and an accumulation-mode design. A depletion-mode design requires a filled two-dimensional electron gas (2DEG) and depletes the QD regions to the few electron regime and has been studied for both doped \cite{Bluhm11,Baart2016,Flentje2017,Payette2012,Prance2012} and undoped \cite{Watson2018,Yoneda2018,Hollmann2020,Struck2020} heterostructures. In the later case, the lack of doping of the 2DEG is compensated by a global top-gate, to which a positive voltage is applied for accumulating charge carriers in the 2DEG. Therefore, on the one hand the gate voltage applied to this global top-gate determines the electron reservoir accumulation, but at the same time it influences the potential in the QD region. This lack of tuning flexibility is often problematic. The problem is solved by an accumulation-mode layout: This is a device with multiple pattern gate-layers, for which positive and negative voltages for accumulation and depletion are chosen within layers \cite{Veldhorst2014,Zajac2018,McJunkin2021, Liu2021,Seidler22}. 

\begin{figure}
    \includegraphics[width=\linewidth]{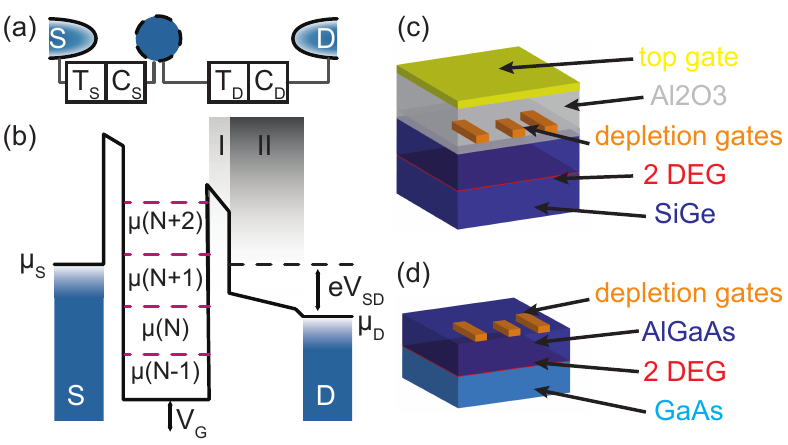}
    \caption{Parameter Settings. \textbf{a} Targeted coupling of a QD (blue) to source (S) and drain (D) reservoir. The tunnel couplings are identical, T$_S$=T$_D$, and the capacitive coupling of the dot to the drain is significantly smaller than to the source, C$_D\ll$C$_S$.  \textbf{b} Schematic potential matching the requirements in panel a. The tunnel coupling and capacitive decoupling is defined by the potential in section I and II, respectively. \textbf{c,d} Simulation input parameterization for Si/SiGe and GaAs/AlGaAs heterostructures, respectively.}
    \label{fig:setup}
\end{figure}

We built a finite-element model of a doped GaAs/AlGaAs and undoped Si/SiGe heterostructures with simplified layer stacks in regards to the permittivity of the different materials (Fig. \ref{fig:setup}c,d) as the basis of the electrostatic simulations. For the GaAs/AlGaAs devices, the depletion gate layer is added directly on top of the heterostructure as metal surfaces. The Si/SiGe devices include an oxide layer underneath the depletion gate layer and between metal gate layers. For the multi-gate layer accumulation design, oxide layers and metal gate layers are alternated. The quantum wells are implemented as a two-dimensional layer within which charges can be accumulated according to the Thomas-Fermi approximation (see Appendix \ref{sec:TFA}) \cite{thomas1927,fermi1928}. We perform the electrostatic simulation using the finite-element solver COMSOL Multiphysics (simulation parameters in Appendix \ref{sec:Parameters}) and determine tunnel-barriers by Wentzel-Kramers-Brillouin (WKB) approximation.

\begin{figure*}
    \includegraphics[width=\textwidth]{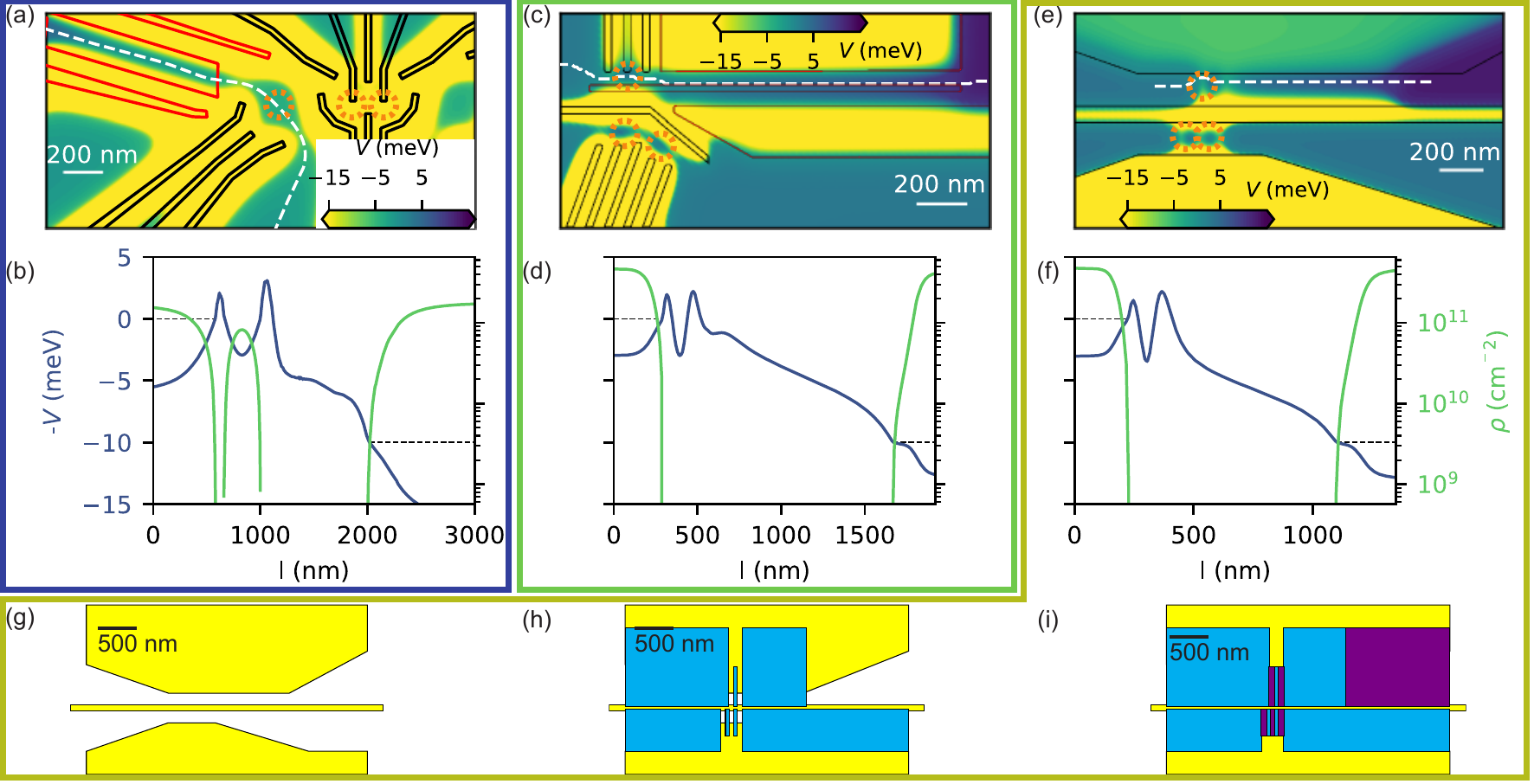}
    \caption{Modeled electrostatic potential. \textbf{a} Potential realization for a doped GaAs/AlGaAs heterostructure. The gate structure (black) and the optimal path (white) are indicated as well as the sensor QD and DQD regions (dashed orange).  \textbf{b} Line cut of potential $V$ (blue) and charge carrier density $\rho$ (green) along the optimal path for the realization in panel a. The source and drain levels are indicated (black dashed). \textbf{c} Potential realization for an undoped Si/SiGe heterostructure with a global top gate. \textbf{d} Line cut of potential and charge carrier density along the optimal path for the realization in panel b. \textbf{e} Potential realization for a closed-gate design in an undoped Si/SiGe heterostructure with the first gate layer (black). \textbf{g-i} Stacking of layers one (yellow), two (blue) and three (purple) of closed gate design.}
    \label{fig1}
\end{figure*}

%Figure 2
We realize the potential for three different boundary conditions: (I) a doped GaAs/AlGaAs heterostructure (II) an undoped Si/SiGe heterostructure with a global topgate and (III) an undoped Si/SiGe heterostructure with an accumulation mode design. Next to the sensor, a DQD is formed in the QW by  properly shaped metal gates. 
For the GaAs/AlGaAs implementation (I, Fig.\ref{fig1}a) the targeted potential is formed by three additional gates added to the sensor design (marked red in Fig. \ref{fig1}a). The angle and shape of which are adjusted in the simulation. In the slide region the potential is depleted without generating a barrier. As the optimal path is nontrivial, the Dijkstra’s algorithm (see Appendix \ref{sec:Dijkstra}) is used to find the optimal path through the tunnel barriers of the sensor and the slide potential (Fig.\ref{fig1}b). The charge density shows that due to the applied bias $V_{\mathrm{SD}}=10\,$mV the 2DEG in the slide region is depleted. 

One Si/SiGe implementation (II, Fig.\ref{fig1}c) uses similarly to the previous case a gate on top of the slide and two side gates (marked red in Fig. \ref{fig1}c). All three gates are set to positive voltages. The side gates have a small angle $\alpha$ between each other to ensure a widening potential. The potential line cut through the sensor and corresponding charge density show the targeted slide potential depleted from charge carriers according to the negligibly small charge density $\rho$ (Fig.\ref{fig1}d).

The second Si/SiGe implementation uses an angle in the confinement gates and a separated accumulation gate in the slide region to define the potential (III, Fig. \ref{fig1}e,f), as shown in the gate structure separated into metal layers (Fig. \ref{fig1}g-i). These simulations allow us to optimize the gate layout without fabricating each design while adhering to the fabrication limitations. For each realization, we are able to judge the feasibility of tuning the voltages applied in each layout to generate the targeted potential. Especially, regions with nontrivial cross-coupling from individual gates (mainly non-orthogonal gate structures) can be studied and optimized to fit a specific potential. Therefore, problems such as required unrealistic fine tuning of specific gates can be avoided.

\section{Benchmark simulations by experiments}
\subsection{Capacitive coupling to sensor QD}
After modelling these very different design layouts, it is indispensable to confirm our simulation methods. To ratify our simulations, we fabricate and test the last design iteration. While the potential is the obvious parameter extracted from the simulations, direct probing of the potential shape for the entire device is hardly possible. Instead we limit ourselves to benchmarking specific measurable properties related to the targeted potential.  As an alternative measure, we choose the capacitive coupling of the  drain reservoir $C_{\mathrm{D}}$ to the sensor QD for this specific application, which can be extracted from both the simulations and transport measurements. $C_{\mathrm{D}}$ allows us to indirectly probe parameters of the specific underlying potential landscape. Simulations of other device layouts can require an adjusted probing measure, with possibilities being calculated and measured tunnel couplings or relative lever arms of different gates among others. $C_{\mathrm{D}}$ can be extracted from the simulations, by varying the bias V$_{\mathrm{SD}}$ applied between the reservoirs and extracting the leverarm of one reservoir on the QD potential (see Appendix \ref{sec:Capacity}).

On the experimental side, measuring a QD in transport formed in the targeted at potential (cf. Fig. \ref{fig:setup}) leads to strongly tilted Coulomb diamonds \cite{Kammerloher2021} from which $C_{\mathrm{D}}$ can be measured. A large and continuously declining slide region decouples the drain reservoir. This decoupling leads to a large negative Coulomb diamond slope $m$, while the positive slope ($m_+$) remains nearly constant.

We correlate the slope $m$ to the capacitive coupling C$_{\mathrm{D}}$ between the drain reservoir and the QD, via the coupling asymmetry $\eta$ to source and drain reservoir: 
\begin{equation}
\begin{split}
    \eta=\left| \frac{dV_{\mathrm{SD}}/dV_{\mathrm{PS}} \nwarrow }{dV_{\mathrm{SD}}/dV_{\mathrm{PS}} \nearrow} \right| = \left| \frac{m }{m_+} \right| \\
   \quad = \left| \frac{C_{\mathrm{PS}}}{C_{\mathrm{D}}}  \cdot \frac{C_{\mathrm{\Sigma}} -C_{\mathrm{D}}}{ C_{\mathrm{PS}}}   \right| =  \left| \frac{C_{\mathrm{\Sigma}} }{C_{\mathrm{D}}} - 1 \right|,
\end{split}
\end{equation}
where $V_{\mathrm{PS}}$ is the voltage applied to the QD plunger PS, $C_{\mathrm{PS}}$ and $C_{\mathrm{\Sigma}}$ are the capacitive coupling of the gate PS and the total capacitive coupling of all gates, respectively.
To probe the potential, we vary the applied gate voltages in the slide region and extract the slopes of the Coulomb diamonds measured for various voltage configurations. For the case II, Si/SiGe with a global accumulation gate, we measure this experimental dependence.

\begin{figure}
    \includegraphics[width=\linewidth]{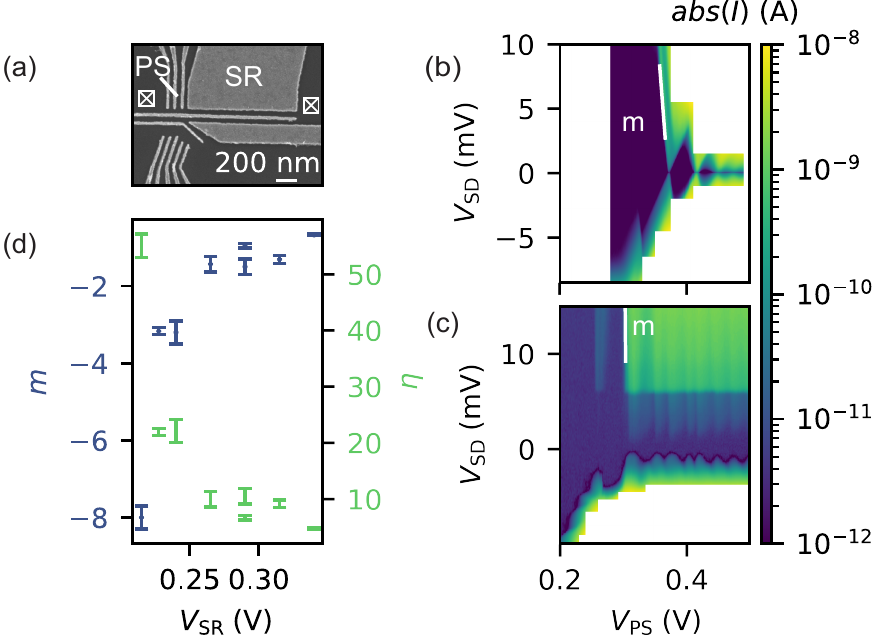}
    \caption{Experimental slide variation. \textbf{a} Scanning electron micrograph (SEM) of an identical device with marked slide gate (SR), plunger (PS) and sensor dot Ohmic contacts (crosses). 
              \textbf{b,c} Coulomb diamonds with marked slope $m$ for $V_{\mathrm{SR}}=0.35$\,V and $V_{\mathrm{SR}}=0.24$\,V, respectively.
              \textbf{d} Coulomb diamond slopes (blue) and $\eta$ (green) in dependence of slide gate voltage.}
    \label{fig:SigeDAta}
\end{figure}

A device similar to the tested one is depicted in Fig. \ref{fig:SigeDAta}a. The voltage of the marked gate SR is adapted to change the potential of the sensor region and control the coupling to the drain reservoir, while the gate PS and the bias applied between the ohmics V$_{\mathrm{SD}}$ are used to measure the diamonds. For larger values of $V_{\mathrm{SR}}$, the standard Coulomb diamonds (Fig.\ref{fig:SigeDAta}b) are measured and the negative slope $m$ of the diamond is extracted. The diamond tilt is enhanced for a decreased voltage applied to gate SR. A steep negative Coulomb diamond slope $m$ is observed for a very low voltage $V_{\mathrm{SR}}$ (Fig. \ref{fig:SigeDAta}c). Comparing $m$ for different slide configurations reveals a systematic decrease in $m$ as a function of a decreasing voltage applied to SR (Fig. \ref{fig:SigeDAta}d). This corresponds to a formation of a slide region after the second barrier which is elongated as lower SR values narrow the path to the drain contact. 

The device layout is optimized for an $\eta=200$ by simulation. Note that larger asymmetries parameterized by $\eta$ are possible according to the simulation, but are of limited use as the corresponding tuning of the plunger voltage applied to gate PS needs to be as accurate as $\Delta V_{\mathrm{PS}}$=0.5\,mV to use the QD as a sensor. The maximum value obtained experimentally during a separated measurement on second sample is $\eta$=85. 
The discrepancy between the simulation and experiments can be attributed to imperfect gate structures or defects as both required an adjustment of the applied voltages, which will shorten the length of the slide region.

Via this indirect probe, we can measure a few aspects of the potential, such as verifying the tunnel barriers and the formation of a tunable slide region. Even without a general or global potential probe, we can ratify our simulation assumptions. 

\subsection{Influence of reservoir bias}
Essential for accurately modelling the potential landscape is not only a description of the gate layer influence, but also a good description of the impact of the electron reservoirs and the applied bias $V_{\mathrm{SD}}$ between the source and the drain reservoir of the sensor. The extend of the reservoirs determines the reach of the screening effects of accumulated electrons. A change in $V_{\mathrm{SD}}$ has a large influence on the overall potential landscape and changes the size of reservoirs. For complex gate layouts and bias $V_{\mathrm{SD}}$ electrostatic landscapes, the extend of the reservoirs is non trivial and can only be predicted by simulations.

\begin{figure*}
    \includegraphics[width=\textwidth]{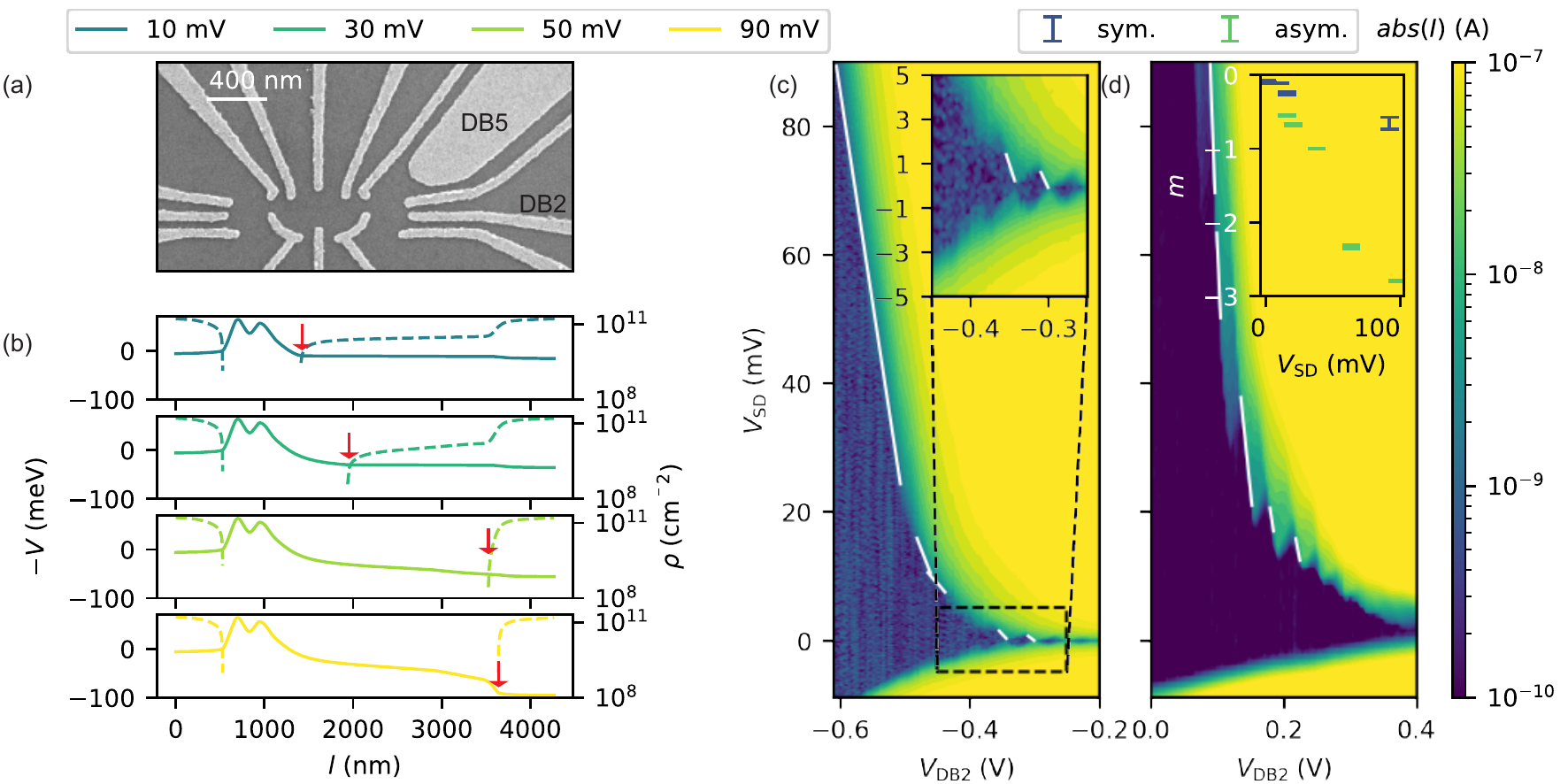}% swith to change panel b
    \caption{Bias Influence. \textbf{a} SEM of an identical device with marked slide gate (DB5) and sensor plunger (DB2).
              \textbf{b} Simulation of bias influence on the formed potential. Only increasing the bias voltage, the slide length of the induced potential $V$ (solid line) increases. Accordingly, the charge density $\rho$ (dashed lines) extends into the slide region depending the applied bias. The beginning of the drain reservoir is marked with red arrows. 
              \textbf{c} Coulomb diamonds of SET operation with asymmetrically applied bias voltage. The slopes (white lines) of the Coulomb diamond are extracted in the dependence of the applied bias. A current level of 1.5\,nA is used as a threshold. For clarity, the low bias regime is depicted as an inset. To realize the SET operation, the slide forming gate voltage $V_{\mathrm{DB5}}$ is set to zero. \textbf{d} Coulomb diamonds of ASD operation with $V_{\mathrm{DB5}}$=-0.31\,V. For the same current threshold $I=1.5$\,nA, the slopes $m$(white lines) are extracted. The extracted $m$ for both operation modes are depicted in the inset in dependence of the applied bias. 
             }
    \label{fig:gaas}
\end{figure*}

To study the bias $V_{\mathrm{SD}}$ influence, we exemplary use a device similar to case I (Fig. \ref{fig1}a), a doped GaAs/AlGaAs device, which was optimized for larger bias values. A device identical to the one measured and simulated is depicted in Fig. \ref{fig:gaas}a.  For this layout, simulations with varying bias voltages V$_{\mathrm{SD}}$ are performed without changing the gate voltages in between (Fig. \ref{fig:gaas}b). The bias $V_{\mathrm{SD}}$ is applied asymmetrically, meaning that only one reservoir potential is shifted. The length of the slide region is determined by the applied $V_{\mathrm{SD}}$ as the reservoir moves into the slide region for low V$_{\mathrm{SD}}$. The boundaries of the reservoir can be identified by the charge carrier density.
We experimentally verify this simulation of the bias impact by comparing to the measured Coulomb diamonds of the device. As discussed for Fig. \ref{fig:SigeDAta}, the extracted capacitive coupling $C_{\mathrm{D}}$ is used as the probe for the potential shape.  For large bias voltages V$_{\mathrm{SD}}$, the Coulomb diamonds tilt even for a QD, which has nearly equal capacitive coupling between sensor QD and its source and drain reservoirs, respectively. We name this type of SET, which does not employ a slide potential, a symmetrically-coupled QD. Here, the asymmetry arise due to a large bias $V_{\mathrm{SD}}$ applied only to one of the two reservoirs. As this device is based on a doped heterostructure, measuring the device with no voltage applied to the gate DB5 allows for the use as a symmetrically-coupled QD. We extract the slope $m$ of the Coulomb diamonds as a function of the  bias $V_{\mathrm{SD}}$ (Fig. \ref{fig:gaas}c). The absolute value of the slope increases with the applied bias $V_{\mathrm{SD}}$. 

An asymmetrically coupled QD can be formed, when depleting the slide region by the voltage applied to gate DB5, forming a potential as intended in the simulations (cf. Fig.\ref{fig:setup}b). Then, steeper diamond slopes are observed (Fig. \ref{fig:gaas}d).  With an added slide region to the potential, the increase of the slope's $m$ absolute value with an increasing bias $V_{\mathrm{SD}}$ is significantly larger, as expected from the simulations, where the slide length and therefore the capacitive coupling of the drain reservoir to the QD strongly depends on the value of the V$_{\mathrm{SD}}$ applied (inset Fig. \ref{fig:gaas}d).
The experimentally observed larger tilt of the Coulomb-diamonds (Fig. \ref{fig:gaas}c vs d) is correctly predicted by our simulation.

\subsection{Fabricational variability}
\begin{figure*}
    \includegraphics[width=\textwidth]{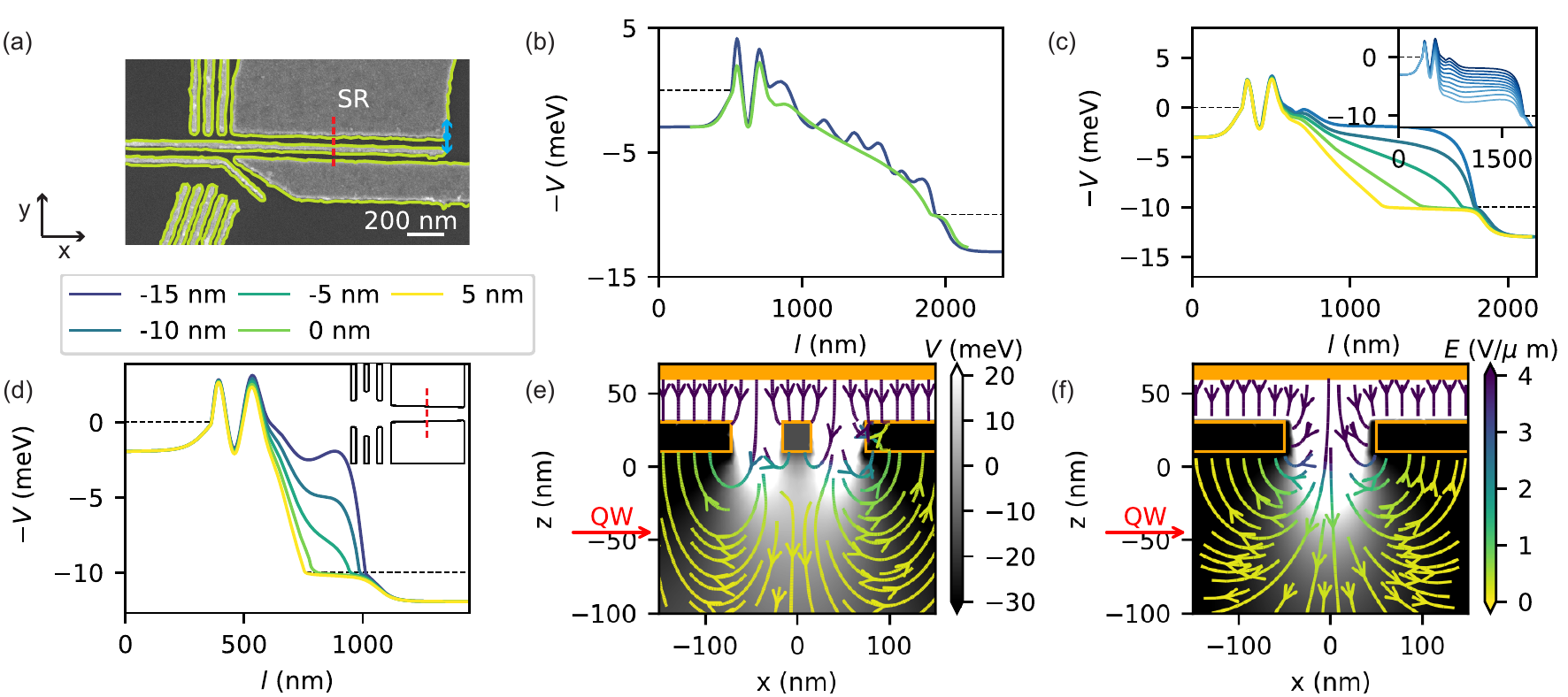}
    \caption{Potential robustness. \textbf{a} SEM of an indentical device. The edges of the fabricated gates (yellow) are extracted by image processing. 
              \textbf{b} Potential line cut along optimal path using the gate edges (panel a) of a fabricated device (blue) and optimal gate edges (green).%( Does it make sense to add the original curve?).
              \textbf{c} Influence of slide angle on optimal path potential. The slide angle is varied by changing the y position of the slide endpoint (blue dot panel a). The inset shows the compensation by slide voltage tuning for $\Delta y= -10$\,nm. Starting from the top $\Delta V_{\mathrm{SR},i}= (i-1)\cdot 10$\,mV. \textbf{d} Slide angle influence for a simpler device (layout in inset). \textbf{e} Cross section of slide potential and electrical field perpendicular to the path at the position marked in red in panel a. The gate edges are marked in orange and the quantum well (red arrow) is placed at z=-45\,nm. \textbf{f} Simple device cross section of slide potential and electrical field perpendicular to the path at the position marked in red in the inset of panel d.
             }
    \label{fig:fig4}
\end{figure*}

A significant benefit of simulating different layouts is the possibility to find a robust design with regard to fabricational influences and imperfections. Therefore, we analyze the stability of the simulated gate-layouts with regard to the spatial resolution limit of the nano-lithography of the metal gates. Exemplary for the geometry of device type II (cf. Fig.\ref{fig1}c), we first study the impact of the line-edge roughness of a patterned gate structure.  The edges of the metallic gates of an exemplary device are identified by applying shape recognition on a scanning-electron micrograph (Fig. \ref{fig:fig4}a). The observed realistic line-edge roughness is fed back to our finite-element device model to predict its impact on the targeted potential shape. The calculated potential along the optimal path predicts a slide potential with multiple ripples after fine tuning the voltages applied when using the realistic line-edge roughness in contrast to the perfect gate edges (Fig. \ref{fig:fig4}b). Although the general potential is obtained, we note that a shorter slide region than compared to the ideal gate-layout is likely to occur as the voltages are tuned to form only one QD with two sharp barriers.

Note that in general it is even not required to fabricate an exemplary device. A simple variation of the shape of the metal gates in the simulation is sufficient to study the robustness to fabrications imperfections. In addition to line-edge roughness \cite{Brauns18}, limitations to accurate alignments of nano-lithography \cite{MOERS12} can be explored by simulation. Non-orthogonal and especially small angled gate structures are not fabricated accurately when relying on electron beam lithography \cite{Brackmann19,RANGELOw94}.  To quantify its impact on the generated potential, we implement small change in angles of the gates in our finite-element model. As an example of angle variation, the position of the corner of the gate labelled SR is changed along the $y$-axis. This results in variations of the opening angle of the gap between the gates in the slide region (bright blue dot Fig. \ref{fig:fig4}a). For a few nanometers’ displacement, our potential simulation predicts that the potential slide is reduced to half its original length or that a flat potential region might occur (yellow and blue line in Fig. \ref{fig:fig4}c, respectively). In the first case, a significantly lower asymmetry $\eta$ would be experimentally obtained causing a decrease in sensor gain \cite{Kammerloher2021}. In the second case the sensor current might be blocked, since the potential disorder (cf. Fig. \ref{fig:fig4}b) alters the flat potential region into a series of disordered QDs and possibly block the current through the sensor region.  

Tuning the voltages applied to specific gates can partly improve the shape of the potential. Also this tunability can be predicted by our device model: For the case of -10\,nm displacement, the tunability of the slide with respect to the $V_{\mathrm{SR}}$ voltage is shown in the inset of Fig. \ref{fig:fig4}c. Tuning the voltage dominantly alters the height of the potential within the slide region and therefore has a small influence on the linear slope of the slide. Our discussions show that gate-layouts, which are unreliable with respect to fabrications imperfections, can be studied and the impact of these imperfections can be eliminated by simulation.

We conclude with a final example of simulation-supported gate design, which has motivated our choice to use the thin metal gate in the device type II (Fig. \ref{fig:fig4}a). For this purpose, we compare a summarized and simplified Si/SiGe device-layout (inset Fig. \ref{fig:fig4}d) of a sensor without a capacitivly coupled DQD. Similar to device type II, we include a global accumulation gate in our device model. The simplification to a symmetric single dot without a DQD nearby allows us to reduce the input parameters. Electrostatic simulation of the variation of the angle of the slide forming gates (cf. Fig. \ref{fig:fig4}c) reveal a larger influence on the slide potential shape (Fig. \ref{fig:fig4}d): A second unintentional potential minimum emerges in the optimal path.

The larger influence of variations in the simple device than the case II device (Fig. \ref{fig1}c) can be understood by our simulations as well: It is caused by the electric field gradients at the quantum well (Figs. \ref{fig:fig4}e,f). For the device type II, the simulated potential forms a single minimum at the height of the quantum well (grey scale Fig. \ref{fig:fig4}e, where a little center gate is included). However, the electric field is mainly defined by the gates patterned on the depletion gate layer while the accumulation gate influence is blocked by the central gate along the current path (colorscaled arrows Fig. \ref{fig:fig4}e). For the simple device without the central thin gate, the influence of the global accumulation gate dominates (Fig \ref{fig:fig4}f). Both the electric field (color scaled arrows) and the potential (greyscale) strongly depend on the global accumulation gate.  For both devices the gates in the depletion gate layer are tuned to have similar voltages to be less sensitive to variations. Thus, we conclude that the thin long gate in device layout II better screens the effect from the top-gate, which has larger potential difference compared to the difference of the slide gates within the depletion layer. For any aimed at design, the consideration of the generated electric fields is useful, as they indicate layout positions which are most impacted by small variations. As it is not possible to fully avoid fabricational fluctuations, reducing generated electric fields improves the robustness and leads to a higher yield. 

\section{Conclusions and outlook}
We show a general approach to electrostatic modelling of spin qubit devices. Our model includes descriptions of the heterostructure, gate layers, reservoirs and applied bias and can be applied to accumulation or depletion type heterostructures. We show gate-designing by device simulation for an exemplary targeted potential shape, which is rather demanding.  We experimentally benchmark our simulation by probing predicted gate-voltage dependencies of current through the device. By simulating our device layout, we were able to predict the properties of and successfully operate first generation devices \cite{Kammerloher2021}. We included a study to make the gate layout robust to unavoidable fabrications imperfections. With our general finite-element modelling of qubit devices, the electrostatic potential landscape can be predicted and the gate-layout optimized without the need of resource-costly fabrication iterations. 

\section*{Acknowledgements}
This work was funded by ARO under the contract NO W911NF-17-1-0349 titled "A scalable and high performance approach to readout of silicon qubits" and by the German Research Foundation (DFG) within the project 289786932 (BO 3140/4-1 and SCHR 1404/2-1).
The device fabrication has been done at HNF - Helmholtz Nano Facility, Research Center Juelich GmbH \cite{Albrecht2017}.

\appendix
\section{Thomas-Fermi approximation}
\label{sec:TFA}
The 2DEG is implemented in the simulation as a two dimensional plane throughout which the charge density $\rho$ determined with the Thomas-Fermi Approximation (TFA):
\begin{equation}
  n_{\mathrm{el}}(x)=\int D(E) f([[E+eV(x)]-\mu]/k_{\mathrm{B}} T)\, d E
\end{equation}
In the case of SiGe, the valley degeneracy leads to a partially defined electron density:
\begin{equation}
    n_{\mathrm{el}}(x)=\left\{\begin{array}{ll}
    \frac{2m^*}{\pi\hbar^2} , & E_{\mathrm{F}}+E_{\mathrm{VS}}<eV(x) \\
    \frac{m^*}{\pi\hbar^2}, & E_{\mathrm{F}}<eV(x)<E_{\mathrm{F}}+E_{\mathrm{VS}} \\
    0 & eV(x)<E_{\mathrm{F}} \\
\end{array}
\right.
\end{equation}
with $E_{\mathrm{F}}$ corresponding to the Fermi energy, $E_{\mathrm{VS}}$ to the valley splitting energy and $V(x)$ to the potential. Since $\rho(V)=e n_{\mathrm{el}}(V)$ and $V=V(\rho)$, $\rho$ and $V$ have to be solved self-consistently. 
To include the applied bias, $V_{\mathrm{SD}}$ is added to $V(x)$, where applicable. The boundaries of these regions are determined by the maximum of the tunnel barriers, where the charge density $\rho=0$.

As the TFA is most accurate for large electron numbers, the QD regions need to be considered carefully. 
For the SiGe simulations, the charge density is set to zero in the QD regions limited by the tunnel barriers. For GaAs, the charge density is measured by TFA also in the QD region.

\section{Simulation parameters}
\label{sec:Parameters}
The parameters used to model the Si/SiGe and AlGaAs/GaAs implementations are listed in Tab. \ref{tab:parameters}. Both the AlGaAs/GaAs and  the Si/SiGe implementation use a fine mesh in the 2DEG region and specifically the QD regions as well as a limited maximum element size to accurately describe the effect of the small gates.
\begin{table}
\begin{center}
\begin{tabular}{|c|c|c|c|}
\hline
    Description & Variable & Si/SiGe & AlGaAs/GaAs \\\hline\hline
    effective  & \multirow{2}{*}{$m^*$} & \multirow{2}{*}{0.19 $m^*_e$}& \multirow{2}{*}{0.067 $m^*_e$}  \\
    electron mass & & &  \\\hline
    valley splitting &$E_{\mathrm{VS}}$ & 70\,$\mu$eV& n.a.  \\\hline
    Fermi energy& $E_F$& 555\,meV& 6.5\,meV\\\hline
    permittivity of  &\multirow{2}{*}{$\epsilon_r$} & \multirow{2}{*}{13} & \multirow{2}{*}{13} \\
    heterostructure & &  &  \\\hline
    permittivity  of oxide &$\epsilon_{\mathrm{oxide}}$ & 11.3 & n.a. \\\hline
    gate height & $h_{\mathrm{gate}}$ & 20\,nm &  30\,nm\\\hline
    oxide height &$h_{\mathrm{oxide}}$ & 10\,nm & n.a. \\\hline
    min. element  &\multirow{2}{*}{$d_{\mathrm{min}}$} & \multirow{2}{*}{5\,nm} & \multirow{2}{*}{1\,nm} \\
    size (ES) &&& \\\hline
    max. ES &$d_{\mathrm{max}}$ & 75\,nm &  150\,nm\\\hline
    min. 2DEG ES &$d_{\mathrm{min,2DEG}}$ & 1\,nm &   1\,nm\\\hline
    max. 2DEG ES &$d_{\mathrm{max,2DEG}}$ & 15\,nm & 15\,nm \\\hline
    depth of 2DEG &$z_{\mathrm{2DEG}}$ & 45\,nm & 90\,nm \\\hline
\end{tabular}    

\caption{COMSOL simulation parameters.}
\label{tab:parameters}
\end{center}
\end{table}

\section{Dijkstra's algorithm for determination of linecuts}
\label{sec:Dijkstra}
As the simulated potential channel has no inherent symmetry, \textit{Dijkstra's algorithm}is used to calculate a well defined path for a linecut through the two-dimensional (2D) potential landscape. This algorithm calculates the cheapest path between two nodes in a graph, where the nodes were defined as the mesh points used for the simulation. The cost function $C$ between the nodes was defined as
\begin{equation}
C = \operatorname{Re}\left(\sqrt{E_F-eV(\textbf{x})}\right)- \epsilon  e V(\textbf{x}),
\end{equation}
where the first part ($\sqrt{E_F-eV(\textbf{x})}$) was based on the semiclassical Wentzel-Kramers-Brillouin (WKB) approximation and the second part ($eV(\textbf{x})$) was based on a classical path-of-lowest-potential. The WKB approximation was thereby used to determine the path through the barriers. Only the real part of the square root was therefore of interest and constants were neglected. The dynamically calculated prefactor $\epsilon$ was chosen to be small, so that the cost function was completely dominated by the WKB contribution and the potential only had an influence outside the barrier zones. \\
The calculated path supports visualization of the relevant potential region. While the well defined cost function allows the comparison between different devices, it does not represent the quantum mechanical behavior of the electron.

\section{Capacity extraction from simulation}
\label{sec:Capacity}
\begin{figure}[ht!]
	\centering
		\includegraphics{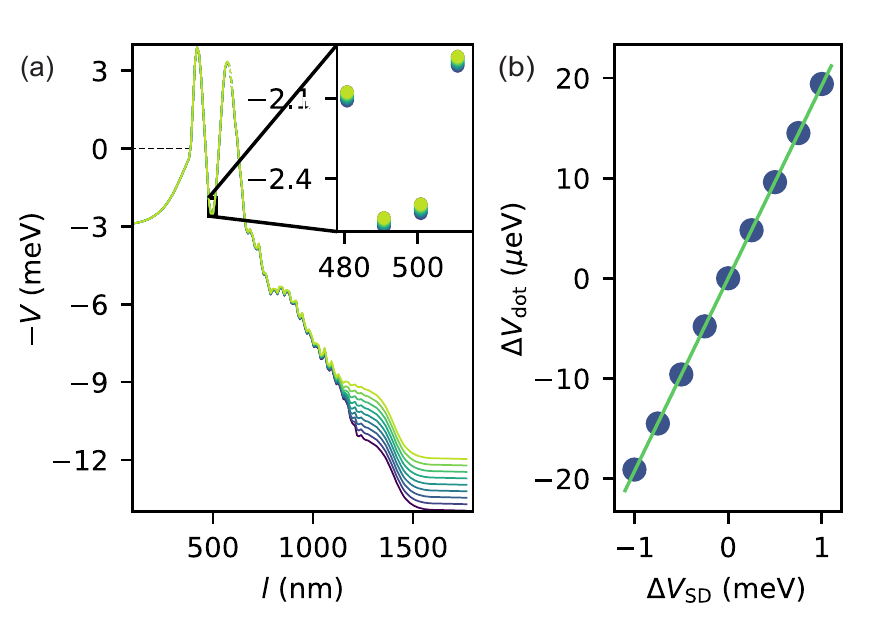} % for other options try v1 or v3
	\caption{Determination of the gate lever-arm $\alpha_{\mathrm{D}}$ by simulation. \textbf{a} Potential line cut along optimal path for bias voltages $V_{\mathrm{SD}}$. \textbf{b} Change in potential minima $\Delta V_{\mathrm{dot}}$, obtained from an interpolation in the QD region  of the 2D data of the potential linecuts in panel a, as a function of the change in bias voltage $\Delta V_{\mathrm{SD}}$. The data points are linearly fitted to obtain the slope $\alpha_{\mathrm{D}}$ (green).}
	\label{fig:AlphaDSimu}
\end{figure}

The coupling asymmetry $\eta$ depends on the capacitance ratio $C_{\mathrm{\Sigma}} C_{\mathrm{D}}^{-1}$. The potential of the dot depends linearly on the voltage applied to the drain reservoir $V_{\mathrm{SD}}$ with the factor $\alpha_{\mathrm{D}} =  - C_{\mathrm{D}} C_{\mathrm{\Sigma}} ^{-1}$.  This was extracted from the simulations by varying $V_{\mathrm{SD}}$ and monitoring the potential minimum $V_0$. $\eta$ is then given by
\begin{equation}
\eta = \left| \frac{C_{\mathrm{\Sigma}} }{C_{\mathrm{D}}} - 1 \right| = \frac{1}{\alpha_{\mathrm{D}}} + 1
\end{equation}
This was realized by running the simulation for a specific design nine times with varying $V_{\mathrm{SD}}$ = -10$\,$mV + [-1$\,$mV, -0.75$\,$mV, -0.5$\,$mV, -0.25$\,$mV, 0$\,$mV, +0.25$\,$mV, +0.5$\,$mV, +0.75$\,$mV, +1$\,$mV]. The potential minimum was determined for each simulation by placing an ellipse around the dot and evaluating the potential on each mesh point inside. As the potential was only evaluated on the points of the discrete mesh, the element density was set accordingly to ensure a high enough resolution to cover possible shifts in the spatial position of the potential minimum. The so obtained potential minima were then plotted against $V_{\mathrm{SD}}$ and fitted to obtain $\alpha_{\mathrm{D}}$ as shown in Fig. \ref{fig:AlphaDSimu}. 

%\section*{Data availability}
%The data sets generated and/or analyzed during this study are available from the corresponding author upon reasonable request.

%\section*{Competing interests}
%The authors declare no competing interests.

%\section*{Author contributions}
%I.S. and M.N. mainly wrote the manuscript. Numerical simulations, the device design and the sample fabrication for Si/SiGe was performed by I.S., M.N. and L.R.S.
%Experiments on the Si/SiGe samples were conducted by A.S., L.D., and D.B., and D.B. grew the Si/SiGe heterostructures.
%The AlGaAs heterostructures were grown by J.R., A.L. and A.W. and devices were fabricated by M.K. For AlGaAs devices, E.K. conducted the experiments and the %numerical simulations advised by H.B. All authors discussed the results and contributed to the manuscript. The study was conceived by L.R.S., H.B. and D.B.

%\bibliography{citations} % Create the reference section using BibTeX

%

\end{document}